\documentclass[prc,superscriptaddress,noshowpacs,unsortedaddress,twocolumn,showpacs,preprintnumbers,amsmath,amssymb]{revtex4-2}

\usepackage[dvipdfmx]{graphicx}
\usepackage{amsmath,amssymb,times}
\usepackage{color}
\usepackage{ulem}
\usepackage{bm}
\usepackage{here}

\newcommand{\beq}{\begin{equation}}
\newcommand{\eeq}{\end{equation}}
\newcommand{\bea}{\begin{eqnarray}}
\newcommand{\eea}{\end{eqnarray}}

\newcommand{\bfi}[1]{\mbox{\boldmath $#1$}}

\newcommand{\vrr}{{\bfi r}}
\newcommand{\vR}{{\bfi R}}

\def\a{\alpha}
\def\b{\beta}

\begin{document}
\title{Determination of matter radius  and neutron-skin thickness of $^{60,62,64}$Ni 
\\ 
from  reaction cross section of proton scattering on $^{60,62,64}$Ni targets}

\author{Shingo~Tagami}
\affiliation{Department of Physics, Kyushu University, Fukuoka 819-0395, Japan}

\author{Tomotsugu~Wakasa}
\affiliation{Department of Physics, Kyushu University, Fukuoka 819-0395, Japan}

\author{Masanobu Yahiro}
\email[]{orion093g@gmail.com}
\affiliation{Department of Physics, Kyushu University, Fukuoka 819-0395, Japan}             

\date{\today}

\begin{abstract}
\noindent 
{\bf Background:}
In our previous work, we determined matter radii $r_{\rm m}({\rm exp})$ 
and neutron-skin thickness 
$r_{\rm skin}({\rm exp})$ from reaction cross sections $\sigma_{\rm R}({\rm exp})$ of proton scattering on $^{208}$Pb, $^{58}$Ni,  $^{40,48}$Ca, 
$^{12}$C targets, using the chiral  (Kyushu) $g$-matrix folding model with 
the densities calculated with Gogny-D1S-HFB (D1S-GHFB) with 
angular momentum  projection (AMP). The resultant $r_{\rm skin}({\rm exp})$ agree 
with the PREX2 and CREX values. 
As for $^{58}$Ni, our value is consistent with one determined from the differential cross section for $^{58}$Ni+$^{4}$He scattering.
As for p+$^{60,62,64}$N scattering, 
$\sigma_{\rm R}({\rm exp})$ are available as a function of incident energies $E_{\rm in}$, 
where $E_{\rm in}=22.8 \sim 65.5$~MeV for $^{60}$Ni, $E_{\rm in}=40,60.8$~MeV for $^{62}$Ni, $E_{\rm in}=40, 60.8$~MeV for $^{64}$Ni.   
\\
{\bf Purpose:} ,
Our aim is to determine matter radii $r_{\rm m}({\rm exp})$ for 
$^{60,62,64}$Ni  from the $\sigma_{\rm R}({\rm exp})$.  
 \\
{\bf Method:} 
Our method is the Kyushu $g$-matrix folding model with the densities scaled from D1S-GHFB+AMP densities, 
\\
{\bf Results:}  
Our skin values are
$r_{\rm skin}({\rm exp})=0.076 \pm	0.019,~0.106 \pm 0.192,~0.162 \pm	0.176$~fm, 
and 
$r_{\rm m}({\rm exp})=3.759 \pm 0.011,~3.811 \pm 0.107,~3.864 \pm	0.101$~fm 
  for $^{60,62,64}$Ni, respectively.    
\end{abstract}

\maketitle


\section{Introduction and conclusion}
\label{Introduction}

\subsection{Background}
{\it Background:}
A novel method for measuring nuclear reactions in inverse kinematics with stored ion beams was successfully used to extract the matter radius $r_{\rm m}({\rm exp})$ 
of $^{58}$Ni~\cite{Zamora:2017adt}. The experiment was performed at the experimental heavy-ion storage ring at the GSI facility. 
Their result determined from the differential cross section 
for $^{58}$Ni+$^{4}$He scattering is $r_m({\rm GSI})=3.70(7)$~fm.

Reaction cross section $\sigma_{\rm R}$ is a standard observable to determine 
$r_{\rm m}({\rm exp})$ and neutron-skin thickness $r_{\rm skin}({\rm exp})$; 
note that 
$r_{\rm skin}$ can be evaluated from the $r_{\rm m}$
by using the $r_{\rm p}({\rm exp})$  calculated
with the isotope shift method based on the electron scattering~\cite{Angeli:2013epw}. 
In fact, we determined $r_{\rm m}({\rm exp})$ and $r_{\rm skin}({\rm exp})$ from 
$\sigma_{\rm R}({\rm exp})$ 
of proton scattering on $^{208}$Pb, $^{58}$Ni,  $^{40,48}$Ca, $^{12}$C targets, 
using the chiral  (Kyushu) $g$-matrix folding model with 
the proton and neutron densities scaled 
with D1S-GHFB+AMP densities~\cite{PhysRevC.107.024608}, 
where D1S-GHFB+AMP is the abbreviation of Gogny-D1S-HFB 
with angular momentum  projection (AMP). 
Our skin values $r_{\rm skin}({\rm exp})$ agree 
with the PREX2 and CREX values. 
As for $^{58}$Ni, our matter radius  
$r_m({\rm exp})=3.711 \pm 0.010$~fm.  
is  consistent with $r_m({\rm GSI})=3.70(7)$~fm. 
As for p+$^{60,62,64}$N scattering, 
$\sigma_{\rm R}({\rm exp})$ are available as a function of incident energy $E_{\rm in}$\cite{INGEMARSSON1999341,PhysRevC.51.1295,PhysRevC.4.1114}; 
here $E_{\rm in}=22.8 \sim 65.5$~MeV for $^{60}$Ni, $E_{\rm in}=40,60.8$~MeV for $^{62}$Ni, $E_{\rm in}=40, 60.8$~MeV for $^{64}$Ni.   
Now we consider p+$^{60,62,64}$N  scattering. 

{\it Purpose:}
Our aim is to determine  $r_{\rm m}({\rm exp})$, 
$r_{\rm skin}({\rm exp})$ for $^{60,62,64}$Ni  from 
the $\sigma_{\rm R}({\rm exp})$ by using the Kyushu $g$-matrix folding model with the proton and neutron densities scaled from D1S-GHFB+AMP ones. 

{\it Results:}
Our results $r_{\rm m}({\rm exp})$, $r_{\rm n}({\rm exp})$, $r_{\rm skin}({\rm exp})$ 
are listed in Table \ref{TW values}.

{\it Conclusion} 
We discover the fact the $r_{\rm m}({\rm exp})/A^{1/3}$ (the matter radius per nucleon) is inversely  proportional to the total binding energy per nucleon $E_{\rm B}/A$, 
where $A$ is the mass number. 

\begin{table}[htb]
\begin{center}
\caption
{Values of   $r_{\rm m}$,  $r_{\rm n}$, $r_{\rm skin}$, $r_{\rm p}$.  
The $r_{\rm p}({\rm exp})$ are determined from the charge radii~\cite{Angeli:2013epw}. 
`Data' shows citations on $\sigma_{\rm R}$. 
The radii are shown in units of fm.  
 }
\begin{tabular}{cccccc}
\hline\hline
  $r_{\rm p}({\rm exp})$ & $r_{\rm m}({\rm exp})$ &  $r_{\rm n}({\rm exp})$ & $r_{\rm skin}({\rm exp})$ & Data \\
\hline
  $^{58}$Ni~  $3.6849$ & $3.711 \pm 0.010 $ & $3.740 \pm 0.019$ & $0.055 \pm 0.019$ & \cite{PhysRevC.107.024608} \\
  $^{60}$Ni~  $3.723$ & $3.759 \pm	0.011$ & $3.799 \pm 0.019$ & $0.076 \pm 0.019$ & \cite{INGEMARSSON1999341,PhysRevC.51.1295} \\
  $^{62}$Ni~  $3.753$ & $3.811 	\pm 0.107$ & $3.859 \pm 0.192$ & $0.106 \pm 0.192$ & \cite{PhysRevC.4.1114}\\
  $^{64}$Ni~  $3.772$ & $3.864 	\pm 0.101$ & $3.933 \pm 0.176$ & $0.162 \pm 0.176$ & \cite{PhysRevC.4.1114}\\
\hline
\end{tabular}
 \label{TW values}
 \end{center} 
 \end{table}

\section{Folding Model}
\label{Sec-Framework}

Kohno calculated the $g$ matrix  for the symmetric nuclear matter, 
using the Brueckner-Hartree-Fock method with chiral N$^{3}$LO 2NFs and NNLO 3NFs~\cite{Kohno:2012vj}. 
He set $c_D=-2.5$ and $c_E=0.25$ so that  the energy per nucleon can  become minimum 
at $\rho = \rho_{0}$. 
Toyokawa {\it et al.} localized the non-local chiral  $g$ matrix~\cite{Toyokawa:2017pdd}, 
using the localization procedure proposed 
by the Melbourne group~\cite{von-Geramb-1991,Amos-1994}. 
The resulting local  $g$ matrix is referred to as  ``local Kyushu  $g$-matrix''.

We use the Kyushu $g$-matrix  folding model~\cite{Toyokawa:2017pdd} 
with the densities calculated with D1S-GHFB+AMP~\cite{Tagami:2019svt}.  
The Kyushu $g$-matrix itself~\cite{Toyokawa:2017pdd}  is constructed from the chiral nucleon-nucleon (NN) interaction with the cutoff 550~MeV.  

In this paper, we consider proton-nucleus scattering. The potential $U(\vR)$ consists of 
the direct and exchange parts,
$U^{\rm DR}(\vR)$ and $U^{\rm EX}(\vR)$~\cite{Minomo:2009ds,Watanabe:2014zea}. 
The validity of the localization is shown in Refs.~\cite{Minomo:2009ds}.

\subsection{Scaling procedure of proton and neutron densities}

For example, the neutron density $\rho_n(r)$ is scaled from the D1S-GHFB+AMP one. 
We can obtain the scaled density $\rho_{\rm scaling}(\vrr)$ from the original density $\rho(\vrr)$ as
\bea
\rho_{\rm scaling}(\vrr)=\frac{1}{\a^3}\rho(\vrr/\a)
\eea
with a scaling factor
\bea
\a=\sqrt{ \frac{\langle \vrr^2 \rangle_{\rm scaling}}{\langle \vrr^2 \rangle}} .
\eea

We scale the neutron density so that  the 
$f \times \sigma_{\rm R}({\rm D1S})$ may reproduce the data ($\sigma_{\rm R}({\rm exp})$) 
under that condition that the $r_{\rm p}({\rm scaling})$ agrees with $r_{\rm p}({\rm exp})$~\cite{Angeli:2013epw} of electron scattering, 
where $\sigma_{\rm R}({\rm D1S})$ is the result of  D1S-GHFB+AMP for each 
$E_{\rm in}$. and $f$ is the average of 
$\sigma_{\rm R}({\rm exp})/\sigma_{\rm R}({\rm D1S})$ over $E_{\rm in}$. 
The matter radius $r_{\rm m}(E_{\rm in})$ thus obtained depends on $E_{\rm in}$. 
We then take the average of $r_{\rm m}(E_{\rm in})$ over  $E_{\rm in}$. 
The resulting value  $r_{\rm m}({\rm exp})$ is shown in Table  \ref{TW values}. 
The corresponding $r_{\rm n}({\rm exp})$ and $r_{\rm skin}({\rm exp})$ 
obtained from $r_{\rm m}({\rm exp})$ and $r_{\rm p}({\rm exp})$ are also shown 
in Table  \ref{TW values}. 

D1M~\cite{Goriely:2009zz,Robledo:2018cdj} is an improved version of D1S 
for binding energies of many nuclei. 
We can use D1M instead of D1S, leading to the change of $U(\vR)$. 
The results of D1M are the same as  our results of Table~ \ref{TW values}.

This scaling procedure is used for proton scattering on Sn
in Ref.~\cite{TAGAMI2023106296}.

\section{Results}
\label{Results}

\begin{figure}[H]
\begin{center}
 \includegraphics[width=0.452\textwidth,clip]{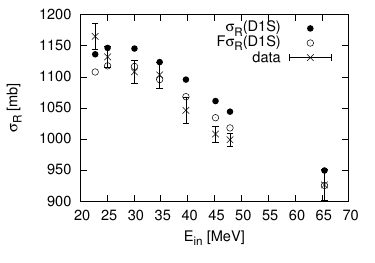}
 \caption{ 
 $E_{\rm in}$ dependence of reaction cross sections $\sigma_{\rm R}$ 
 for $p$+$^{60}$Ni scattering. 
 Closed circles denote $\sigma_{\rm R}({\rm D1S})$ of the  D1S-GHFB+AMP densities, 
 whereas  open circles correspond  to $f \times \sigma_{\rm R}({\rm D1S})$. 
 The data (crosses) are taken from 
 Refs.~\cite{INGEMARSSON1999341,PhysRevC.51.1295}.
   }
 \label{Fig-RXsec-p+Ni60}
\end{center}
\end{figure}

 Figure \ref{Fig-RXsec-p+Ni60} shows reaction cross sections $\sigma_{\rm R}$ 
 as a function of $E_{\rm in}$.
Our result $\sigma_{\rm R}({\rm D1S})$ overshoots somewhat, but 
$f \times \sigma_{\rm R}({\rm D1S})$ is close to the central values of experimental 
data~\cite{INGEMARSSON1999341,PhysRevC.51.1295}, where $f=0.97488$. 
The $f \times \sigma_{\rm R}({\rm D1S})$ are used in order to determine 
$r_{\rm m}({\rm exp})$ and $r_{\rm skin}({\rm exp})$.

The same scaling procedure is taken also for $p$+$^{62,64}$Ni  scattering, 
where $F=0.96213,~0.98232$ for $^{62,64}$Ni, respectively.

The data on $E_{\rm B}/A$ (the total binding energy per nucleon) 
weakly depend on $A$ for $^{58,60,62,64}$Ni~\cite{HP:NuDat2.8}. This is true for 
$r_{\rm m}({\rm exp})/A^{1/3}$ (the matter radius per nucleon). 
In fact, $A$ dependence of  $r_{\rm m}({\rm exp})/A^{1/3} \times E_{\rm B}/A$ is even smaller; namely, 
the average $\b$ of $r_{\rm m}({\rm exp})/A^{1/3} \times E_{\rm B}/A$ over 
$A$ is 
\bea
\b=8.437385~ {\rm MeV\cdot fm}
\label{Eq-a}
\eea
and the error is $0.024~ {\rm MeV\cdot fm}$. 
The relative error is 0.2847\%. This indicates that 
 the $r_{\rm m}({\rm exp})/A^{1/3}$ (the matter radius per nucleon) is in inverse  proportional to the total binding energy per nucleon $E_{\rm B}/A$; 
 see Table \ref{TW values-2} for $E_{\rm B}/A$ and $r_{\rm m}({\rm exp})/A^{1/3}$. 

Using $\b=8.437385~ {\rm MeV\cdot fm}$, 
$E_{\rm B}/A$, we can derive the central values of 
$r_{\rm m}({\rm exp})$. These are 
3.711, 
3.759, 
3.811, 
3.864 
~fm 
for $^{58,60,62,64}$Ni. 
These values agree with our values of Table \ref{TW values}. 
One can easily evaluate the central value of matter radius   
from $\b$, $E_{\rm B}/A$. This is convenient.

\begin{table}[htb]
\begin{center}
\caption
{Values of  $E_{\rm B}/A $~\cite{HP:NuDat2.8}, $r_{\rm m}({\rm exp})/A^(1/3)$. 
The $r_{\rm m}({\rm exp})$ are shown in units of fm and the  $E_{\rm B}/A $ are in 
MeV.
 }
\begin{tabular}{cccccc}
\hline\hline
  $BE/A $ & $r_{\rm m}({\rm exp})/A^(1/3)$  \\
\hline
  $^{58}$Ni~  $8.732059$ & $0.959$  \\
  $^{60}$Ni~  $8.780774$ & $0.960$  \\
  $^{62}$Ni~  $8.794553$ & $0.963$ \\
  $^{64}$Ni~  $8.777461$ & $0.966$ \\
\hline
\end{tabular}
 \label{TW values-2}
 \end{center} 
 \end{table}

\noindent
\appendix

\noindent
\begin{acknowledgments}
We would like to thank Dr. Toyokawa for his contribution. 
\end{acknowledgments}


\bibliography{Folding-v18}

\begin{thebibliography}{17}%
\makeatletter
\providecommand \@ifxundefined [1]{%
 \@ifx{#1\undefined}
}%
\providecommand \@ifnum [1]{%
 \ifnum #1\expandafter \@firstoftwo
 \else \expandafter \@secondoftwo
 \fi
}%
\providecommand \@ifx [1]{%
 \ifx #1\expandafter \@firstoftwo
 \else \expandafter \@secondoftwo
 \fi
}%
\providecommand \natexlab [1]{#1}%
\providecommand \enquote  [1]{``#1''}%
\providecommand \bibnamefont  [1]{#1}%
\providecommand \bibfnamefont [1]{#1}%
\providecommand \citenamefont [1]{#1}%
\providecommand \href@noop [0]{\@secondoftwo}%
\providecommand \href [0]{\begingroup \@sanitize@url \@href}%
\providecommand \@href[1]{\@@startlink{#1}\@@href}%
\providecommand \@@href[1]{\endgroup#1\@@endlink}%
\providecommand \@sanitize@url [0]{\catcode `\\12\catcode `\$12\catcode
  `\&12\catcode `\#12\catcode `\^12\catcode `\_12\catcode `\%12\relax}%
\providecommand \@@startlink[1]{}%
\providecommand \@@endlink[0]{}%
\providecommand \url  [0]{\begingroup\@sanitize@url \@url }%
\providecommand \@url [1]{\endgroup\@href {#1}{\urlprefix }}%
\providecommand \urlprefix  [0]{URL }%
\providecommand \Eprint [0]{\href }%
\providecommand \doibase [0]{https://doi.org/}%
\providecommand \selectlanguage [0]{\@gobble}%
\providecommand \bibinfo  [0]{\@secondoftwo}%
\providecommand \bibfield  [0]{\@secondoftwo}%
\providecommand \translation [1]{[#1]}%
\providecommand \BibitemOpen [0]{}%
\providecommand \bibitemStop [0]{}%
\providecommand \bibitemNoStop [0]{.\EOS\space}%
\providecommand \EOS [0]{\spacefactor3000\relax}%
\providecommand \BibitemShut  [1]{\csname bibitem#1\endcsname}%
\let\auto@bib@innerbib\@empty
\bibitem [{\citenamefont {Zamora}\ \emph {et~al.}(2017)\citenamefont {Zamora}
  \emph {et~al.}}]{Zamora:2017adt}%
  \BibitemOpen
  \bibfield  {author} {\bibinfo {author} {\bibfnamefont {J.~C.}\ \bibnamefont
  {Zamora}} \emph {et~al.},\ }\bibfield  {title} {\bibinfo {title}
  {{Nuclear-matter radius studies from
  Ni58(\ensuremath{\alpha},\ensuremath{\alpha}) experiments at the GSI
  Experimental Storage Ring with the EXL facility}},\ }\href
  {https://doi.org/10.1103/PhysRevC.96.034617} {\bibfield  {journal} {\bibinfo
  {journal} {Phys. Rev. C}\ }\textbf {\bibinfo {volume} {96}},\ \bibinfo
  {pages} {034617} (\bibinfo {year} {2017})}\BibitemShut {NoStop}%
\bibitem [{\citenamefont {Angeli}\ and\ \citenamefont
  {Marinova}(2013)}]{Angeli:2013epw}%
  \BibitemOpen
  \bibfield  {author} {\bibinfo {author} {\bibfnamefont {I.}~\bibnamefont
  {Angeli}}\ and\ \bibinfo {author} {\bibfnamefont {K.~P.}\ \bibnamefont
  {Marinova}},\ }\bibfield  {title} {\bibinfo {title} {{Table of experimental
  nuclear ground state charge radii: An update}},\ }\href
  {https://doi.org/10.1016/j.adt.2011.12.006} {\bibfield  {journal} {\bibinfo
  {journal} {Atom. Data Nucl. Data Tabl.}\ }\textbf {\bibinfo {volume} {99}},\
  \bibinfo {pages} {69} (\bibinfo {year} {2013})}\BibitemShut {NoStop}%
\bibitem [{\citenamefont {Wakasa}\ \emph {et~al.}(2023)\citenamefont {Wakasa},
  \citenamefont {Tagami}, \citenamefont {Matsui}, \citenamefont {Takechi},\
  and\ \citenamefont {Yahiro}}]{PhysRevC.107.024608}%
  \BibitemOpen
  \bibfield  {author} {\bibinfo {author} {\bibfnamefont {T.}~\bibnamefont
  {Wakasa}}, \bibinfo {author} {\bibfnamefont {S.}~\bibnamefont {Tagami}},
  \bibinfo {author} {\bibfnamefont {J.}~\bibnamefont {Matsui}}, \bibinfo
  {author} {\bibfnamefont {M.}~\bibnamefont {Takechi}},\ and\ \bibinfo {author}
  {\bibfnamefont {M.}~\bibnamefont {Yahiro}},\ }\bibfield  {title} {\bibinfo
  {title} {Neutron-skin values and matter and neutron radii determined from
  reaction cross sections of proton scattering on $^{12}\mathrm{C}$,
  $^{40,48}\mathrm{Ca}$, $^{58}\mathrm{Ni}$, and $^{208}\mathrm{Pb}$},\ }\href
  {https://doi.org/10.1103/PhysRevC.107.024608} {\bibfield  {journal} {\bibinfo
   {journal} {Phys. Rev. C}\ }\textbf {\bibinfo {volume} {107}},\ \bibinfo
  {pages} {024608} (\bibinfo {year} {2023})}\BibitemShut {NoStop}%
\bibitem [{\citenamefont {Ingemarsson}\ \emph {et~al.}(1999)\citenamefont
  {Ingemarsson}, \citenamefont {Nyberg}, \citenamefont {Renberg}, \citenamefont
  {Sundberg}, \citenamefont {Carlson}, \citenamefont {Auce}, \citenamefont
  {Johansson}, \citenamefont {Tibell}, \citenamefont {Clark}, \citenamefont
  {{Kurth Kerr}},\ and\ \citenamefont {Hama}}]{INGEMARSSON1999341}%
  \BibitemOpen
  \bibfield  {author} {\bibinfo {author} {\bibfnamefont {A.}~\bibnamefont
  {Ingemarsson}}, \bibinfo {author} {\bibfnamefont {J.}~\bibnamefont {Nyberg}},
  \bibinfo {author} {\bibfnamefont {P.}~\bibnamefont {Renberg}}, \bibinfo
  {author} {\bibfnamefont {O.}~\bibnamefont {Sundberg}}, \bibinfo {author}
  {\bibfnamefont {R.}~\bibnamefont {Carlson}}, \bibinfo {author} {\bibfnamefont
  {A.}~\bibnamefont {Auce}}, \bibinfo {author} {\bibfnamefont {R.}~\bibnamefont
  {Johansson}}, \bibinfo {author} {\bibfnamefont {G.}~\bibnamefont {Tibell}},
  \bibinfo {author} {\bibfnamefont {B.}~\bibnamefont {Clark}}, \bibinfo
  {author} {\bibfnamefont {L.}~\bibnamefont {{Kurth Kerr}}},\ and\ \bibinfo
  {author} {\bibfnamefont {S.}~\bibnamefont {Hama}},\ }\href
  {https://doi.org/https://doi.org/10.1016/S0375-9474(99)00236-5} {\bibfield
  {journal} {\bibinfo  {journal} {Nuclear Physics A}\ }\textbf {\bibinfo
  {volume} {653}},\ \bibinfo {pages} {341} (\bibinfo {year}
  {1999})}\BibitemShut {NoStop}%
\bibitem [{\citenamefont {Eliyakut-Roshko}\ \emph {et~al.}(1995)\citenamefont
  {Eliyakut-Roshko}, \citenamefont {McCamis}, \citenamefont {van Oers},
  \citenamefont {Carlson},\ and\ \citenamefont {Cox}}]{PhysRevC.51.1295}%
  \BibitemOpen
  \bibfield  {author} {\bibinfo {author} {\bibfnamefont {T.}~\bibnamefont
  {Eliyakut-Roshko}}, \bibinfo {author} {\bibfnamefont {R.~H.}\ \bibnamefont
  {McCamis}}, \bibinfo {author} {\bibfnamefont {W.~T.~H.}\ \bibnamefont {van
  Oers}}, \bibinfo {author} {\bibfnamefont {R.~F.}\ \bibnamefont {Carlson}},\
  and\ \bibinfo {author} {\bibfnamefont {A.~J.}\ \bibnamefont {Cox}},\
  }\bibfield  {title} {\bibinfo {title} {Measurements of proton total reaction
  cross sections for $^{58}\mathrm{Ni}$ and $^{60}\mathrm{Ni}$ including
  nonrelativistic and relativistic data analyses},\ }\href
  {https://doi.org/10.1103/PhysRevC.51.1295} {\bibfield  {journal} {\bibinfo
  {journal} {Phys. Rev. C}\ }\textbf {\bibinfo {volume} {51}},\ \bibinfo
  {pages} {1295} (\bibinfo {year} {1995})}\BibitemShut {NoStop}%
\bibitem [{\citenamefont {Menet}\ \emph {et~al.}(1971)\citenamefont {Menet},
  \citenamefont {Gross}, \citenamefont {Malanify},\ and\ \citenamefont
  {Zucker}}]{PhysRevC.4.1114}%
  \BibitemOpen
  \bibfield  {author} {\bibinfo {author} {\bibfnamefont {J.~J.~H.}\
  \bibnamefont {Menet}}, \bibinfo {author} {\bibfnamefont {E.~E.}\ \bibnamefont
  {Gross}}, \bibinfo {author} {\bibfnamefont {J.~J.}\ \bibnamefont
  {Malanify}},\ and\ \bibinfo {author} {\bibfnamefont {A.}~\bibnamefont
  {Zucker}},\ }\bibfield  {title} {\bibinfo {title}
  {Total-reaction-cross-section measurements for 30-60-mev protons and the
  imaginary optical potential},\ }\href
  {https://doi.org/10.1103/PhysRevC.4.1114} {\bibfield  {journal} {\bibinfo
  {journal} {Phys. Rev. C}\ }\textbf {\bibinfo {volume} {4}},\ \bibinfo {pages}
  {1114} (\bibinfo {year} {1971})}\BibitemShut {NoStop}%
\bibitem [{\citenamefont {Kohno}(2012)}]{Kohno:2012vj}%
  \BibitemOpen
  \bibfield  {author} {\bibinfo {author} {\bibfnamefont {M.}~\bibnamefont
  {Kohno}},\ }\bibfield  {title} {\bibinfo {title} {{Strength of reduced
  two-body spin-orbit interaction from chiral three-nucleon force}},\ }\href
  {https://doi.org/10.1103/PhysRevC.86.061301} {\bibfield  {journal} {\bibinfo
  {journal} {Phys. Rev. C}\ }\textbf {\bibinfo {volume} {86}},\ \bibinfo
  {pages} {061301} (\bibinfo {year} {2012})},\ \Eprint
  {https://arxiv.org/abs/1209.5048} {arXiv:1209.5048 [nucl-th]} \BibitemShut
  {NoStop}%
\bibitem [{\citenamefont {Toyokawa}\ \emph {et~al.}(2018)\citenamefont
  {Toyokawa}, \citenamefont {Yahiro}, \citenamefont {Matsumoto},\ and\
  \citenamefont {Kohno}}]{Toyokawa:2017pdd}%
  \BibitemOpen
  \bibfield  {author} {\bibinfo {author} {\bibfnamefont {M.}~\bibnamefont
  {Toyokawa}}, \bibinfo {author} {\bibfnamefont {M.}~\bibnamefont {Yahiro}},
  \bibinfo {author} {\bibfnamefont {T.}~\bibnamefont {Matsumoto}},\ and\
  \bibinfo {author} {\bibfnamefont {M.}~\bibnamefont {Kohno}},\ }\bibfield
  {title} {\bibinfo {title} {{Effects of chiral three-nucleon forces on
  $^{4}$He-nucleus scattering in a wide range of incident energies}},\ }\href
  {https://doi.org/10.1093/ptep/pty001} {\bibfield  {journal} {\bibinfo
  {journal} {PTEP}\ }\textbf {\bibinfo {volume} {2018}},\ \bibinfo {pages}
  {023D03} (\bibinfo {year} {2018})},\ \Eprint
  {https://arxiv.org/abs/1712.07033} {arXiv:1712.07033 [nucl-th]} \BibitemShut
  {NoStop}%
\bibitem [{\citenamefont {von Geramb}\ \emph {et~al.}(1991)\citenamefont {von
  Geramb} \emph {et~al.}}]{von-Geramb-1991}%
  \BibitemOpen
  \bibfield  {author} {\bibinfo {author} {\bibfnamefont {H.~V.}\ \bibnamefont
  {von Geramb}} \emph {et~al.},\ }\href@noop {} {\bibfield  {journal} {\bibinfo
   {journal} {Phys. Rev. C}\ }\textbf {\bibinfo {volume} {44}},\ \bibinfo
  {pages} {73} (\bibinfo {year} {1991})}\BibitemShut {NoStop}%
\bibitem [{\citenamefont {Amos}\ and\ \citenamefont
  {Dortmans}(1994)}]{Amos-1994}%
  \BibitemOpen
  \bibfield  {author} {\bibinfo {author} {\bibfnamefont {K.}~\bibnamefont
  {Amos}}\ and\ \bibinfo {author} {\bibfnamefont {P.~J.}\ \bibnamefont
  {Dortmans}},\ }\href@noop {} {\bibfield  {journal} {\bibinfo  {journal}
  {Phys. Rev. C}\ }\textbf {\bibinfo {volume} {49}},\ \bibinfo {pages} {1309}
  (\bibinfo {year} {1994})}\BibitemShut {NoStop}%
\bibitem [{\citenamefont {Tagami}\ \emph {et~al.}(2020)\citenamefont {Tagami},
  \citenamefont {Tanaka}, \citenamefont {Takechi}, \citenamefont {Fukuda},\
  and\ \citenamefont {Yahiro}}]{Tagami:2019svt}%
  \BibitemOpen
  \bibfield  {author} {\bibinfo {author} {\bibfnamefont {S.}~\bibnamefont
  {Tagami}}, \bibinfo {author} {\bibfnamefont {M.}~\bibnamefont {Tanaka}},
  \bibinfo {author} {\bibfnamefont {M.}~\bibnamefont {Takechi}}, \bibinfo
  {author} {\bibfnamefont {M.}~\bibnamefont {Fukuda}},\ and\ \bibinfo {author}
  {\bibfnamefont {M.}~\bibnamefont {Yahiro}},\ }\bibfield  {title} {\bibinfo
  {title} {{Chiral $g$-matrix folding-model approach to reaction cross sections
  for scattering of Ca isotopes on a C target}},\ }\href
  {https://doi.org/10.1103/PhysRevC.101.014620} {\bibfield  {journal} {\bibinfo
   {journal} {Phys. Rev. C}\ }\textbf {\bibinfo {volume} {101}},\ \bibinfo
  {pages} {014620} (\bibinfo {year} {2020})},\ \Eprint
  {https://arxiv.org/abs/1911.05417} {arXiv:1911.05417 [nucl-th]} \BibitemShut
  {NoStop}%
\bibitem [{\citenamefont {Minomo}\ \emph {et~al.}(2010)\citenamefont {Minomo},
  \citenamefont {Ogata}, \citenamefont {Kohno}, \citenamefont {Shimizu},\ and\
  \citenamefont {Yahiro}}]{Minomo:2009ds}%
  \BibitemOpen
  \bibfield  {author} {\bibinfo {author} {\bibfnamefont {K.}~\bibnamefont
  {Minomo}}, \bibinfo {author} {\bibfnamefont {K.}~\bibnamefont {Ogata}},
  \bibinfo {author} {\bibfnamefont {M.}~\bibnamefont {Kohno}}, \bibinfo
  {author} {\bibfnamefont {Y.~R.}\ \bibnamefont {Shimizu}},\ and\ \bibinfo
  {author} {\bibfnamefont {M.}~\bibnamefont {Yahiro}},\ }\bibfield  {title}
  {\bibinfo {title} {{The Brieva-Rook Localization of the Microscopic
  Nucleon-Nucleus Potential}},\ }\href
  {https://doi.org/10.1088/0954-3899/37/8/085011} {\bibfield  {journal}
  {\bibinfo  {journal} {J. Phys. G}\ }\textbf {\bibinfo {volume} {37}},\
  \bibinfo {pages} {085011} (\bibinfo {year} {2010})},\ \Eprint
  {https://arxiv.org/abs/0911.1184} {arXiv:0911.1184 [nucl-th]} \BibitemShut
  {NoStop}%
\bibitem [{\citenamefont {Watanabe}\ \emph {et~al.}(2014)\citenamefont
  {Watanabe} \emph {et~al.}}]{Watanabe:2014zea}%
  \BibitemOpen
  \bibfield  {author} {\bibinfo {author} {\bibfnamefont {S.}~\bibnamefont
  {Watanabe}} \emph {et~al.},\ }\bibfield  {title} {\bibinfo {title}
  {{Ground-state properties of neutron-rich Mg isotopes}},\ }\href
  {https://doi.org/10.1103/PhysRevC.89.044610} {\bibfield  {journal} {\bibinfo
  {journal} {Phys. Rev. C}\ }\textbf {\bibinfo {volume} {89}},\ \bibinfo
  {pages} {044610} (\bibinfo {year} {2014})},\ \Eprint
  {https://arxiv.org/abs/1404.2373} {arXiv:1404.2373 [nucl-th]} \BibitemShut
  {NoStop}%
\bibitem [{\citenamefont {Goriely}\ \emph {et~al.}(2009)\citenamefont
  {Goriely}, \citenamefont {Hilaire}, \citenamefont {Girod},\ and\
  \citenamefont {Peru}}]{Goriely:2009zz}%
  \BibitemOpen
  \bibfield  {author} {\bibinfo {author} {\bibfnamefont {S.}~\bibnamefont
  {Goriely}}, \bibinfo {author} {\bibfnamefont {S.}~\bibnamefont {Hilaire}},
  \bibinfo {author} {\bibfnamefont {M.}~\bibnamefont {Girod}},\ and\ \bibinfo
  {author} {\bibfnamefont {S.}~\bibnamefont {Peru}},\ }\bibfield  {title}
  {\bibinfo {title} {{First Gogny-Hartree-Fock-Bogoliubov Nuclear Mass
  Model}},\ }\href {https://doi.org/10.1103/PhysRevLett.102.242501} {\bibfield
  {journal} {\bibinfo  {journal} {Phys. Rev. Lett.}\ }\textbf {\bibinfo
  {volume} {102}},\ \bibinfo {pages} {242501} (\bibinfo {year}
  {2009})}\BibitemShut {NoStop}%
\bibitem [{\citenamefont {Robledo}\ \emph {et~al.}(2019)\citenamefont
  {Robledo}, \citenamefont {Rodr\'\i{}guez},\ and\ \citenamefont
  {Rodr\'\i{}guez-Guzm\'an}}]{Robledo:2018cdj}%
  \BibitemOpen
  \bibfield  {author} {\bibinfo {author} {\bibfnamefont {L.~M.}\ \bibnamefont
  {Robledo}}, \bibinfo {author} {\bibfnamefont {T.~R.}\ \bibnamefont
  {Rodr\'\i{}guez}},\ and\ \bibinfo {author} {\bibfnamefont {R.~R.}\
  \bibnamefont {Rodr\'\i{}guez-Guzm\'an}},\ }\bibfield  {title} {\bibinfo
  {title} {{Mean field and beyond description of nuclear structure with the
  Gogny force: A review}},\ }\href {https://doi.org/10.1088/1361-6471/aadebd}
  {\bibfield  {journal} {\bibinfo  {journal} {J. Phys. G}\ }\textbf {\bibinfo
  {volume} {46}},\ \bibinfo {pages} {013001} (\bibinfo {year} {2019})},\
  \Eprint {https://arxiv.org/abs/1807.02518} {arXiv:1807.02518 [nucl-th]}
  \BibitemShut {NoStop}%
\bibitem [{\citenamefont {Tagami}\ \emph {et~al.}(2023)\citenamefont {Tagami},
  \citenamefont {Wakasa},\ and\ \citenamefont {Yahiro}}]{TAGAMI2023106296}%
  \BibitemOpen
  \bibfield  {author} {\bibinfo {author} {\bibfnamefont {S.}~\bibnamefont
  {Tagami}}, \bibinfo {author} {\bibfnamefont {T.}~\bibnamefont {Wakasa}},\
  and\ \bibinfo {author} {\bibfnamefont {M.}~\bibnamefont {Yahiro}},\
  }\bibfield  {title} {\bibinfo {title} {Neutron skin thickness of
  116,118,120,122,124sn determined from reaction cross sections of proton
  scattering},\ }\href
  {https://doi.org/https://doi.org/10.1016/j.rinp.2023.106296} {\bibfield
  {journal} {\bibinfo  {journal} {Results in Physics}\ }\textbf {\bibinfo
  {volume} {46}},\ \bibinfo {pages} {106296} (\bibinfo {year}
  {2023})}\BibitemShut {NoStop}%
\bibitem [{HP:()}]{HP:NuDat2.8}%
  \BibitemOpen
  \href@noop {} {}\bibinfo {note}
  {{h}ttps://www.nndc.bnl.gov/nudat2/}\BibitemShut {NoStop}%
\end{thebibliography}%

\end{document}